\documentclass{article}
\usepackage{ijcai18}

\usepackage{times}
\usepackage{xcolor}
\usepackage{soul}
\usepackage[utf8]{inputenc}
\usepackage[small]{caption}
\usepackage{algorithm}
\usepackage{algorithmicx}
\usepackage{algpseudocode}
\usepackage{amsmath}
\usepackage{graphicx}
\usepackage{IEEEtrantools}
\usepackage{multirow}

\title{Globally Optimized Mutual Influence Aware Ranking in E-Commerce Search}

\author{
Tao Zhuang$^1$,
Wenwu Ou$^2$,
Zhirong Wang$^3$ \\ 
Taobao Search, Alibaba Group Holding Limited\\
$^1$zhuangtau@gmail.com,
$^2$santong.oww@taobao.com,
$^3$qingfeng@taobao.com
}
\begin{document}

\maketitle

\begin{abstract}
In web search, mutual influences between documents have been studied from the perspective of search result diversification.
But the methods in web search is not directly applicable to e-commerce search because of their differences.
And little research has been done on the mutual influences between items in e-commerce search.
We propose a global optimization framework for mutual influence aware ranking in e-commerce search.
Our framework directly optimizes the Gross Merchandise Volume (GMV) for ranking, and decomposes ranking into two tasks.
The first task is mutual influence aware purchase probability estimation.
We propose a global feature extension method to incorporate mutual influences into the features of an item.
We also use Recurrent Neural Network (RNN) to capture influences related to ranking orders in purchase probability estimation.
The second task is to find the best ranking order based on the purchase probability estimations.
We treat the second task as a sequence generation problem and solved it using the beam search algorithm.
We performed online A/B test on a large e-commerce search engine.
The results show that our method brings a 5\% increase in GMV for the search engine over a strong baseline.
\end{abstract}

\section{Introduction}
In web search, the importance of mutual influences between documents has been recognized early~\cite{Carbonell:98,Zhai:03}, and has been studied from the perspective of search result diversification~\cite{Radlinski:08,Agrawal:09,Zhu:14,Xia:17}, because relevance and diversity are major concerns of web search.
However, mutual influences between items in e-commerce search are quite different.
When a customer issue a query on a large e-commerce platform, usually thousands of highly relevant items are returned.
The customer has to compare the items and selects the one that best suits his needs.
The customer's comparison is usually not on relevance because all items are highly relevant, but on several detailed aspects of an item, e.g. the price, brand, quality etc.
If an item is surrounded by others with similar quality but much higher prices, then its probability of being purchased would be high.
On the contrary, if the same item is surrounded by items of much lower prices, then its probability of being purchased would be lower.
Therefore, in e-commerce search, mutual influences between items are even stronger than those in web search.
Moreover, the goals and metrics of e-commerce search and web search are different.
The goal of e-commerce search is to help a customer make a purchase.
So the most widely used metric for e-commerce search in industry is the GMV of the search engine, rather than the Normalized Discounted Cumulative Gain (NDCG) for web search.
And search result diversity is not a major concern in e-commerce search.
Because of these differences, the methods of search result diversification in web search are not suitable for e-commerce search.
However, in spite of the importance of mutual influences in e-commerce search ranking, little research has been done in this area.

We propose a framework for mutual influence aware ranking in e-commerce search for the first time.
We formulate ranking as a global optimization problem with the objective to maximize the mathematical expectation of GMV.
In our framework, mutual influences between items are considered in the purchase probability estimation model, which predicts whether an item will be purchased.
We first propose a global feature extension method to incorporate mutual influences into the features of an item.
And we use a DNN to estimate purchase probability based on the extended features.
With our DNN model, the optimal ranking is found by simple sorting.
Furthermore, we use RNN to consider mutual influences related to ranking orders.
And we design a novel attention mechanism for the RNN model to capture long-distance influences between items.
With our RNN model, our global optimization framework becomes a sequence generation problem.
We use the beam search algorithm to generate a good ranking.
We performed online A/B test on Taobao Search\footnote{Taobao Search is the major e-commerce search service provided by Alibaba Group. See: https://s.taobao.com/}, one of the largest e-commerce search engines in the world.
We compared the GMV and computational cost of our methods with a strong baseline.
The results show that with acceptable computational cost, our method brings a 5\% increase in GMV over the baseline, which is regarded as a really big improvement in Taobao Search.

\section{Related Work}
The works on learning to rank in general~\cite{Liu:09,Chapelle:11a} are related to us.
The pointwise~\cite{Cossock:08,Li:08}, pairwise~\cite{Joachims:02,Burges:07}, and listwise approaches~\cite{Cao:07,Xia:08,Xu:07,Taylor:08} score each document {\em individually}, and sort the documents according to their scores.
So mutual influences between documents are not explicitly considered.
Our work explicitly models the mutual influences between items in ranking.

The works on search result diversification are closely related to ours.
Agrawal~{\it et~al.}~\shortcite{Agrawal:09} diversify the search results by assuming the existence of a taxonomy of information.
We do not make such assumptions.
Instead, we use mutual influence aware purchase probability estimation for ranking.
Radlinski~{\it et~al.}~\shortcite{Radlinski:08} use the multi-armed bandit algorithm to explore and learn a diverse ranking of documents.
We find a good ranking using an online optimization framework and do not perform online explorations.
The works in~\cite{Carbonell:98,Zhu:14,Xia:17} treated ranking as a sequential selection process, where the documents are selected step by step.
Carbonell~{\it et~al.}~\shortcite{Carbonell:98} propose to select a new document with maximal additional utility at each step.
Zhu~{\it et~al.}~\shortcite{Zhu:14} score a document using features from the previously selected documents and select the current document using a greedy algorithm.
Xia~{\it et~al.}~\shortcite{Xia:17} model the sequential selection process as a Markov Decision Process and make a greedy selection at each step.
Our method differs from the three works above in that we model the sequential selection process using RNN models and find a ranking using beam search algorithm instead of a greedy algorithm.

The work of Wang~{\it et~al.}~\shortcite{Wang:16} is similar to ours in methodology.
They also use a optimization framework to present search results.
But they focus on how to render heterogeneous results from different sources on the same page, rather than how to rank documents.
We focus on the ranking task and propose a optimization framework with novel ranking models.

\section{Globally Optimized Mutual Influence Aware Reranking}\label{sec:ranking_models}
Let $S = (1,...,N)$ denote the set of items to be ranked.
Let $O$ denote the set of all permutations on $S$, and $o = (o_1, ..., o_N) \in O$ is one permutation.
Let $d_i$ denote the order of item $i$ in $o$, i.e. $o_{d_i} = i$.
For the item $i$, influences from other items are determined by its {\bf ranking context}: $c(o,i) = (o_1, ..., o_{d_i-1}, o_{d_i+1}, ..., o_N)$.
The probability that a customer will purchase item $i$ in ranking $o$ is denoted as $p(i | c(o,i))$.
Let $v(i)$ denote the price of item $i$.
According to the online statistics, when a customer decides to buy an item, the amount he buys is 1 most of the times.
So the sales value of item $i$ is $v(i)$.
Then the expected GMV of the ranking $o$ is: 
\begin{equation}
E(\text{GMV} | o) = \sum_{i=1}^{N} v(i) p(i | c(o,i))
\end{equation}
The goal of ranking is to find a permutation that maximize the expected GMV:
\begin{equation}\label{eqn:optimize}
o^{*} = \arg\max_{o \in O} \sum_{i=1}^{N} v(i) p(i | c(o,i))
\end{equation}
The problem in Equation~(\ref{eqn:optimize}) is decomposed to two problems:
\begin{description}
    \item[Problem 1] How to accurately estimate $p(i | c(o,i))$.
    \item[Problem 2] How to efficiently find $o^*$.
\end{description}

We will refer to these two problems as Problem 1 and 2 in the following.
Problem 1 is purchase probability estimation and is solved using discriminative machine learning methods.
Problem 2 is global optimization with a search space of $N!$ permutations.
And problem 2 needs to be solved in real time online.
Practically, we simplify $c(o,i)$ in reasonable ways to facilitate the solution of Problem 2.
We present two simplifications and solutions in the following.
\subsection{Global Feature Extension}\label{subsec:dnn_model}
In the first simplification, we only consider the set of items in ranking context, i.e. $c(o,i) = S$.
The ranking orders are ignored in $c(o,i)$.
In ranking, item $i$ is represented by a feature vector $f_l(i)$, which we call the local features of $i$.
Usually each dimension of $f_l(i)$ is a real-valued number.
We and we assume that  in this paper.
In this simplification, influences from other items to $i$ are represented by a global feature vector $f_g(i)$.
We concatenate the local and global features of $i$ as $f(i) = (f_l(i), f_g(i))$. and use $f(i)$ to predict $p(i | c(o,i))$.
We call $f(i)$ the {\bf global feature extension} of item $i$.

The global features of an item are generated as follows.
Let us take the price feature as an example.
We can compare the price of $i$ with the prices of others to see whether $i$ is expensive in $S$.
Suppose price is the first dimension of $f_l$, i.e. $f_{l1}(i)$ is the price feature of $i$, then we can do the comparison by computing $f_{g1}(i) = (f_{l1}(i) - f_{\text{min},1}) / (f_{\text{max},1} - f_{\text{min},1})$, where $f_{\text{min},1}$ and $f_{\text{max},1}$ are the lowest and highest prices in $S$ respectively.
Obviously $f_{g1}(i)$ lies between $[0, 1]$ and measures the relative ``expensiveness" of $i$ in $S$.
Similarly, we compare every local feature of $i$ with others in this way and obtain the global feature vector $f_g(i)$:
\begin{IEEEeqnarray}{rCl}
    \label{eqn:feat_min}
    f_{\text{min}} = \min_{1 \le j \le N} f_{l}(j)\\
    \label{eqn:feat_max}
    f_{\text{max}} = \max_{1 \le j \le N} f_{l}(j)\\
    \label{eqn:feat_g}
    f_g(i) = \frac{ f_{l}(i) - f_{\text{min}} }{ f_{\text{max}} - f_{\text{min}} }
\end{IEEEeqnarray}
Note that all variables in Equations~(\ref{eqn:feat_min}-\ref{eqn:feat_g}) are vectors.
So all operators in Equations~(\ref{eqn:feat_min}-\ref{eqn:feat_g}) are applied element-wisely to each dimension of the vectors.
The global features of $i$ measures the relative local feature values of $i$ compared to others in $S$.
Suppose the number of local features is $d$, then the time complexity of global feature extension is $\Theta(Nd)$,
 which is acceptable so long as $d$ is not too large.

\subsubsection{The DNN Model}
The global feature extension $f(i)$ incorporates influences from other items to $i$.
So we feed $f(i)$ to a DNN to make mutual influence aware estimation of $p(i | c(o, i))$.
Our DNN has 3 hidden ReLU layers and a sigmoid output layer and is trained with the cross-entropy loss.
After the global feature extension procedure, $p(i | c(o,i))$ can be calculated for each $i$ independently.
However, the ranking position of $i$ is not considered in this model.
So the position bias~\cite{Craswell:08} will be a problem. 
We remedy this by multiplying a position bias to our model: $bias(d_i) p(i | c(o,i))$, where $d_i$ is the ranking position of $i$. 
Note that position bias is a decreasing function with respect to ranking positions.
So we do not need to know the exact values of position bias, because Problem 2 can be solved by scoring each item with $v(i) p(i | c(o,i))$ and sorting them in descending order.
The proof is simple and omitted here.
Therefore, we get a very efficient ranking method.
In total, $\Theta(Nd)$ time is needed for global feature extension and $\Theta(N)$ forward runs of DNN are needed to find an optimal ranking.
\subsection{Ranking as Sequence Generation}\label{subsec:rnn}
In the second simplification, when estimating $p(i | c(o,i))$, we consider not only the set of items, but also the ranking orders ahead of item $i$, i.e. $c(o,i) = (o_{1,2,...,d_{i}-1}, S)$.
Similar to the first simplification, we use global feature extensions.
Then we can write $p(i | c(o,i)) = p(i | o_{1},o_{2},...,o_{d_{i}-1})$, because $S$ has already been considered by the global feature extensions.
Estimating $p(i | c(o,i))$ based on ranking orders ahead of $i$ conforms to a customer's browsing habit, because a customer usually browses in a top-down manner.
The ranking orders after $i$ are not considered mainly because that will break the sequential selection process, which supports efficient incremental computations as we will show.
\subsubsection{The Basic RNN Model}
Now Problem 1 becomes a sequential probability estimation problem.
And an RNN model is used to estimate $p(i | c(o,i))$.
The estimation probability is computed iteratively as follows:
\begin{IEEEeqnarray}{rCl}\label{eqn:rnn}
    p(o_i | o_{1,..., i-1})&=&\text{sigmoid}(W_h h_i)\\
    h_i&=&\text{RNN}(h_{i-1}, f(o_i))
\end{IEEEeqnarray}
Note that in Equation~(\ref{eqn:rnn}) we use the notation $p(o_i | o_{1,..., i-1})$ instead of $p(i | o_{1},o_{2},...,o_{d_{i}-1})$ for convenience.

Problem 2 now becomes a sequence generation problem, which is similar to the decoding process in machine translation~\cite{Sutskever:14}.
The difference is that we need to maximize the expected GMV and each item in $S$ must appear exactly once in the sequence.
Despite of the difference, the efficient beam search algorithm can still be adapted to solve Problem 2.
Beam search is a generalization of the greedy search algorithm.
Beam search keeps the top-$k$ sequences at each step of selection, and returns the best sequence at the last step.
And $k$ is called the {\bf beam size}.
Compared to greedy algorithm, beam search explores a larger search space and finds a better solution, although it cannot guarantee to find the optimal solution.
The beam search algorithm for ranking is presented in Algorithm~\ref{alg:beamsearch}.
\begin{algorithm}[t!]
  \caption{Beam search for Ranking with RNN model}\label{alg:beamsearch}
  \begin{algorithmic}[1]
      \Require The set of items: $S = \{1, 2, ..., N\}$; The beam size $k$;
  An item feature map: $f$;
  An item value map: $v$;
  An RNN model: $\text{RNN}(h, f)$.
  \Ensure A sequence $s$ with max GMV.
  \State $\text{beam} \gets \text{maxheap}([(0: \langle\phi,0\rangle)])$.
  \State $h \gets 0$; $c \gets 0$.
  \For{$n$ from $1$ to $N$}
    \State stepbeam $\gets$ an empty max-heap.
    \For{\textbf{each} $(c: \langle s, h \rangle)$ in beam}  \Comment{$c$ is the key in the heap.}
      \State candidates $\gets$ an empty max-heap.
      \For{\textbf{each} $i$ in $S \setminus s$}
        \State $h' \gets \text{RNN}(h, f(i))$
        \State $p(i | s) \gets \text{sigmoid}(W_h h')$
        \State $s' \gets \text{append } i \text{ to } s$
        \State $c' \gets c + v(i)p(i | s)$
        \State insert $(c': \langle s', h' \rangle)$ to candidates, keeping size $\le k$.
      \EndFor
      \State merge candidates to stepbeam, keeping size $\le k$.
    \EndFor
    \State beam $\gets$ stepbeam.
  \EndFor
  \State \textbf{return} The top sequence in beam.
  \end{algorithmic}
\end{algorithm}

As shown in Algorithm~\ref{alg:beamsearch}, RNN is very suitable for incremental computations.
As long as we remember the hidden vector $h$ of last step, we can compute the current hidden vector $h' = \text{RNN}(h, f(i))$, and the purchase probability $p(i | c(o,i)) = p(i | s) = \text{sigmoid}(W_h h')$, where $s$ is the sequence of items ahead of item $i$.
And the expected GMV of the sequence $s'$ is $c + v(i)p(i | s)$ where $c$ is the expected GMV of $s$.
These incremental computations are quite efficient.
The time complexity of Algorithm~\ref{alg:beamsearch} is $\Theta(kN^{2})$, where the unit time is that of computing an $\text{RNN}(h,f)$.
Although the quadratic time complexity is usually too expensive for ranking, we can use it in a reranking process where only top-$N$ items are reranked.
Details will be explained in Section~\ref{subsec:online_test}.
For clarity, we point out that beam search is only used in online ranking.
The training of the RNN model does not need to use beam search.

\subsubsection{RNN Model with Attention}
\begin{figure}[!tbh]
    \centering
    \includegraphics[width=0.5\textwidth]{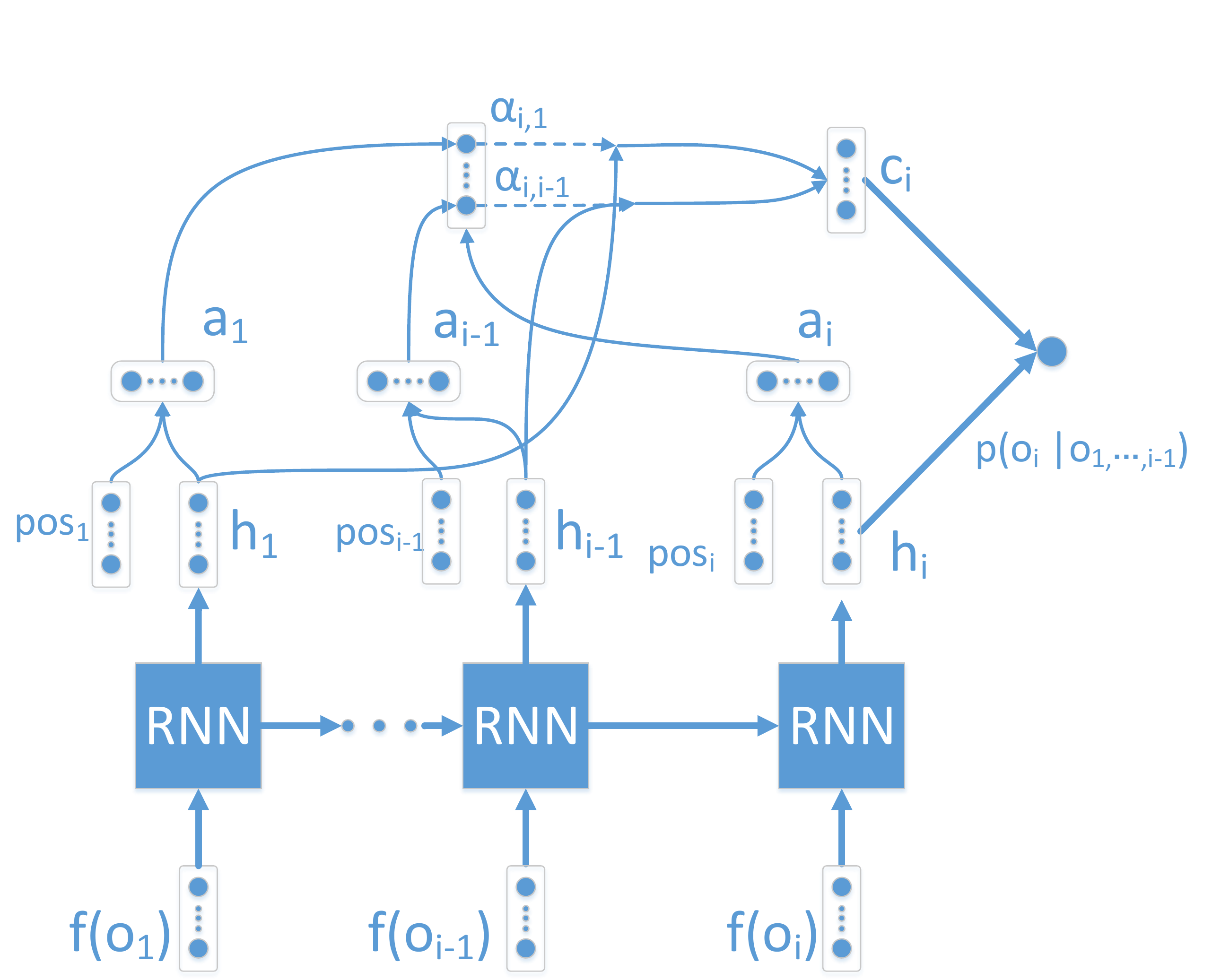}
    \caption{
        {\small The RNN with attention model for purchase probability estimation.}
    \label{fig:attention}}
\end{figure}

We use LSTM as the RNN cell.
But in practice we find that LSTM cannot learn the long-range dependency.
We use RNN to calculate the purchase probability at the 20th position in many sequences and find that $p(o_{20} | o_{1,....19})$, $p(o_{20} | o_{2,....19})$, $p(o_{20} | o_{3,....19})$, $p(o_{20} | o_{4,....19})$ are almost the same.
This means that items ranked at the top 4 positions have little impact on the purchase probability of the item at the 20th position.
This is not reasonable because a customer usually has strong impressions on the top ranked items.
And influences from the top items should not be neglected.
To overcome this problem, we design an attention mechanism to the RNN model, inspired by the work on neural machine translation~\cite{bahdanau:14}.
The network structure is shown in Figure~\ref{fig:attention}.
In our attention mechanism, $p(o_i | o_{1,..., i-1})$ depends not only on current the hidden vector $h_i$, but also on the weighted sum of all previous hidden vectors.
The vector of weights is: $\alpha_i = \alpha_{ij}, 1 \le j < i$, where $\alpha_{ij}$ is the attention of $o_i$ to $h_j$.
\begin{IEEEeqnarray}{rCl}
p(o_i | o_{1,..., i -1})&=&\text{sigmoid}(W_h h_i + W_c c_i)\\
h_i&=&\text{RNN}(h_{i-1}, f(o_i))\\
c_i&=&\sum_{j=1}^{i-1} \alpha_{ij} h_{j}
\end{IEEEeqnarray}
This mechanism ensures that any previous item $o_j, j < i$ has a direct influence to $p(o_i | o_{1,..., i-1})$, so that the long-range influences will not diminish.
$\alpha_{ij}$ is computed as:
\begin{IEEEeqnarray}{rCl}\label{eqn:attention}
    \alpha_{ij}&=&\frac{\exp(g_{ij})}{\sum_{k=1}^{i-1} \exp(g_{ik})}
\end{IEEEeqnarray}
And $g_{ij}$ is the score that measures the influence of $o_j$ to $o_i$, computed as a ReLU layer:
\begin{IEEEeqnarray}{rCl}
g_{ij}&=&\text{ReLU}(W_g \begin{bmatrix}   a_i  \\   a_j \\ \end{bmatrix})
\end{IEEEeqnarray}
$a_i$ is a representaions of $o_i$ and is computed as:
\begin{IEEEeqnarray}{rCl}\label{eqn:ai}
a_i&=&\text{ReLU}(W_a  \begin{bmatrix}   pos_i  \\   h_i \\ \end{bmatrix})
\end{IEEEeqnarray}
$pos_i$ is an embedding vector for ranking position $i$.
All the parameters and position embeddings are learned on training data.

With the RNN attention model, we still use beam search to find a good ranking sequence.
We only need to change the computation of $p(i|s)$ and $h'$ in Algorithm~\ref{alg:beamsearch} from the basic RNN model to the RNN attention model.
But the attention mechanism requires more computation.
And the time complexity of Algorithm~\ref{alg:beamsearch} becomes $\Theta(kN^3)$, where the unit time is that of computing an $\text{RNN}(h,f)$ or an $\alpha_{ij}$ as in Equation~(\ref{eqn:attention}). 
The time complexity increases by a factor of $N$ because at position $i$, the attentions at previous positions $\alpha_{ij}, 1 \le j < i$ need to be computed.

\section{Experiments}\label{sec:experiments}

\subsection{Experimental Setup}
Our experiments are carried out on the Taobao Search platform, which is one of the largest e-commerce search services in the world, with a current daily GMV of billions in CNY.
Our training data for purchase probability estimation are from the query logs of Taobao Search.
Each record in the log corresponds to a query, and contains the items presented to a customer.
The purchased items are positive samples and others are negative samples.
Positive samples are much fewer than negative ones.
To balance the number of positive and negative samples, we discard all records that contain no purchase. 
So in our training data, each record contains at least one purchase.
By doing so, we not only reduce the number of negative samples, but also exclude those records in which the user has no purchase intention and just browses for fun.
In online test, we update our model on a daily basis and use the last day's log data for training.

The local features of an item include the price, the relevance to query, the Click Through Rate (CTR), the Click-purchase Conversion Rate (CVR), and various user preference scores such as brand or shop preferences, etc.
Except the price feature, each local feature is produced by a lower-level model.
And most lower-level models use huge number of lower-level features, e.g. the relevance model uses query words and the headline of an item as features.
Our ranking model is built upon these local features and thus is freed from using huge number of lower-level features.
In fact, we only use 23 local features in our purchase probability model.
We must point out that most local features we use are personalized and {\em real time} features, which means they will change in real time according to a user's feedbacks such as clicking or purchasing items.
The real time features enable our model to utilize a user's real time feedbacks.

In the following, we will refer our DNN, RNN, and RNN with attention models in Section~\ref{sec:ranking_models} as `miDNN', `miRNN', and `miRNN+attention' respectively.
The input feature size is 46.
In miDNN model, the sizes of the three hidden layers are 50, 50, 30.
In RNN models, the hidden vector size of LSTM is 50.
The sizes of $a_i$ and $pos_i$ in Equation~(\ref{eqn:ai}) are 10 and 5 respectively.

\subsubsection{The Baseline Algorithm} 
Our baseline is a 5-layer DNN ranking model that was in online operation in Taobao Search at the time of our experiments.
The baseline uses the same local features with our models, but does not use our global feature extension in Section~\ref{subsec:dnn_model}. 
Therefore, the baseline only uses the local features, whereas our models uses both local featuers and the global feature extension.
The baseline is also trained to estimate the purchase probability of an item.
The ranking score of item $i$ is computed as $v(i)^\gamma p(i)$, where $v(i)$ is the price of $i$ and $p(i)$ is the estimated purchase probability.
Items are ranked in descending scores.
The parameter $\gamma$ is fine-tuned by human through online A/B test to maximize GMV.
This baseline had been fine tuned for a long time and was the best algorithm in online operation.
In the following, we will refer to the baseline algorithm as `DNN'.

\subsection{Evaluation of Purchase Probability Estimations}
We first compare the purchase probability estimation accuracies between our models and the baseline.
We use one day's log data for training and the next day's log data for test.
Both the training and test data contain around 17 million records.
And on average there are about 50 items in each record.
So there are about 850 million items in both training and test data.
For the baseline DNN model each item is a sample.
For our models each record is used as a sequence of items, with each item being a sample.
We use the Area Under the ROC Curve (AUC) and Relative Information Gain (RIG)~\cite{Chen:12} metrics for evaluation.
The results on test data are shown in Table~\ref{tab:prob_est_rst}.
\begin{table}[!htb]
    \centering
    {\small
        \begin{tabular}{|c|c|c|}
            \hline
            Models & AUC & RIG \\
            \hline
            DNN & 0.724 & 0.094 \\
            miDNN & 0.747 & 0.119 \\
            miRNN & 0.765 & 0.141 \\
            miRNN+attention & {\bf 0.774} & ${\bf 0.156}$ \\
            \hline
        \end{tabular}
    }
    \caption{The test results of purchase probability estimation.\label{tab:prob_est_rst}}
\end{table}

\begin{figure}[!tbh]
    \centering
    \includegraphics[width=0.4\textwidth]{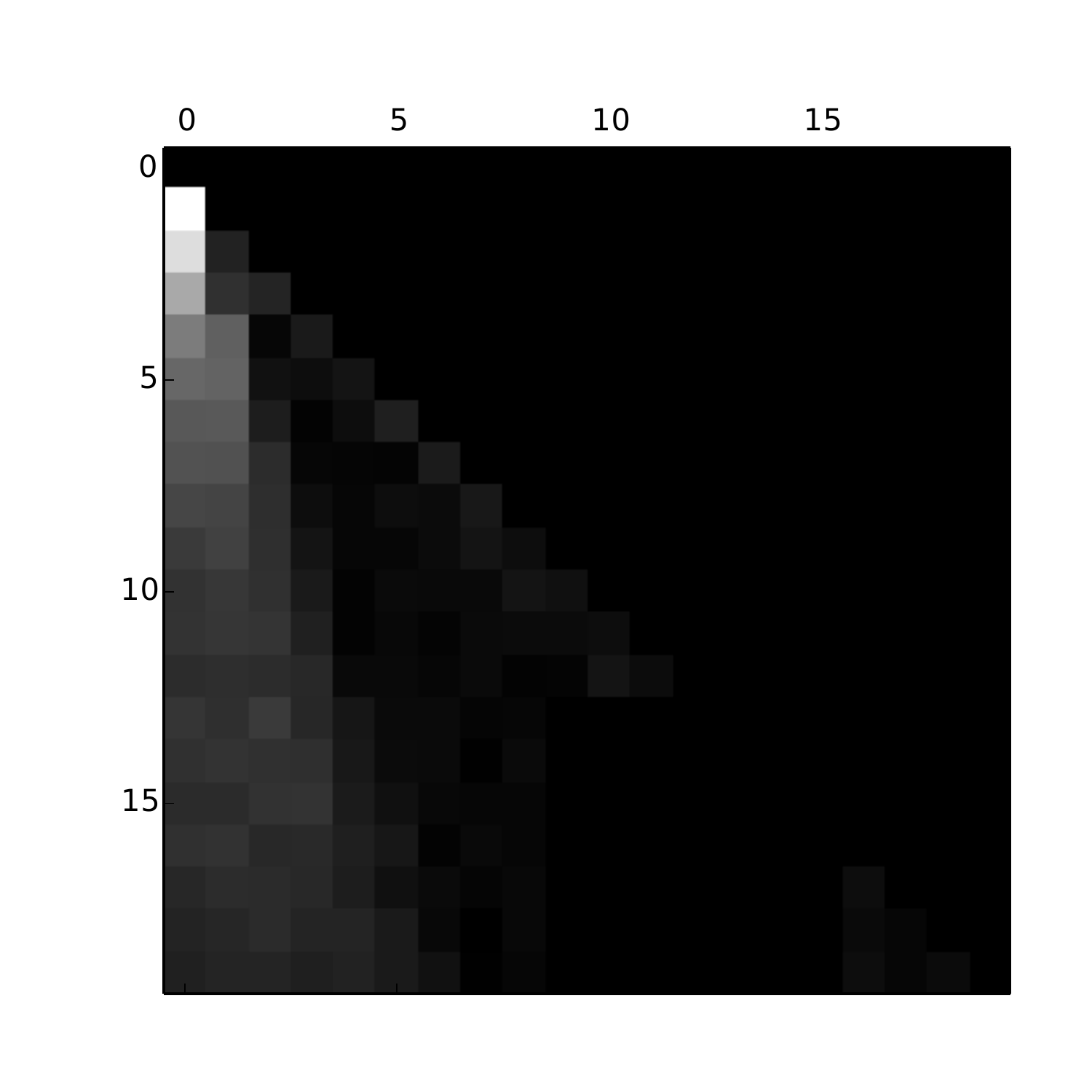}
    \caption{
        {\small The average attention matrix when $N = 20$.}
    \label{fig:attention_mat}}
\end{figure}

From Table~\ref{tab:prob_est_rst} we see that the AUC and RIG of miDNN are much better than the baseline DNN.
Note that the major difference between DNN and miDNN is that miDNN uses the global feature extension in Section~\ref{subsec:dnn_model} whereas DNN does not.
This indicates that the global features are really useful and mutual influences between items are important indeed.
By considering orders, miRNN has better results than miDNN.
The best results are achieved by miRNN+attention, with a noticeable improvement over the results of miRNN.
To see why miRNN+attention has better results, we visualize the attention $\alpha_{ij}$ from Equation~(\ref{eqn:attention}) as follows.
We find all test records whose length is at least 20, and calculate the $\alpha_{ij}, i \le 20, j < i$.
Then we average the values of $\alpha_{ij}$ from different records to obtain $\overline{\alpha}_{ij}$.
So $\overline{\alpha}_{ij}$ is the average attention of position $i$ to position $j$.
The $\overline{\alpha}_{ij}$ for $i \le 20, j < i$ are shown in Figure~\ref{fig:attention_mat}, where each row corresponds to an $i$ and each column to a $j$.

From Figure~\ref{fig:attention_mat} we see the top ranked items have larger attentions even when $i=20$.
This means that our RNN with attention learns to consider the influences from top ranked items even when estimating the probability of an item ranked far below.
So the attention mechanism indeed alleviates the long-distance dependency problem that we stated in Section~\ref{subsec:rnn}.


\subsection{Online A/B Test}\label{subsec:online_test}
We also performed online A/B test to compare the GMV of our models to the baseline.
In online A/B test, users, together with their queries, are randomly and evenly distributed to 30 buckets.
Each experimental algorithm of ours is deployed in 1 bucket.
And the baseline algorithm is deployed in the baseline bucket.
The GMVs of experimental buckets and the baseline bucket are compared.
Our online tests lasted for a time of a month to accumulate reliable results. 

\begin{figure}[!tbh]
    \centering
    \includegraphics[width=0.34\textwidth]{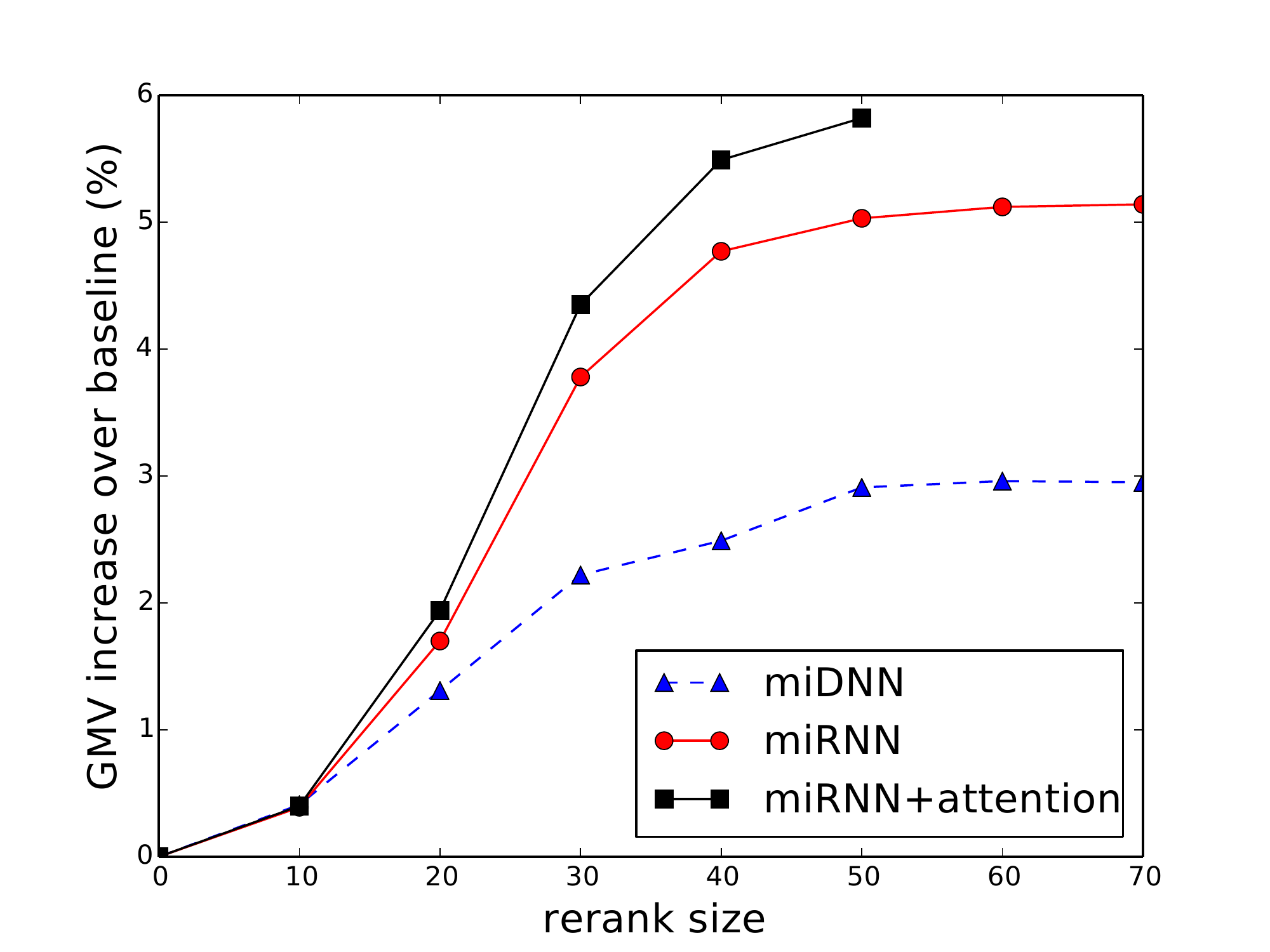}
    \caption{
        {\small The GMV increase with respect to rerank size.}
    \label{fig:gmv_reranksize}}
\end{figure}

For each query, usually thousands of items need to be ranked.
But statistics show that over 92\% users browse no more than 100 items.
And the average number of items browsed by users is 52.
This means that only the top ranked items really matter.
And as we stated in Section~\ref{sec:ranking_models}, the time complexity of miRNN and miRNN+attention are $\Theta(kN^2)$ and $\Theta(kN^3)$.
For practical considerations on computing efficiency, the RNN models should not rank too many items.
Therefore, we use our models in a reranking process in online tests.
Specifically, we rerank the top-$N$ items of baseline ranking using our models.
And we call $N$ the {\bf rerank size} of our models.
We experimented with different rerank sizes and compare the GMVs of our models to the baseline.
The beam sizes of our RNN models are set to 5 unless stated otherwise.
The relative GMV increase of our models over the baseline is shown in Figure~\ref{fig:gmv_reranksize}.

\begin{figure}[!tbh]
    \centering
    \includegraphics[width=0.34\textwidth]{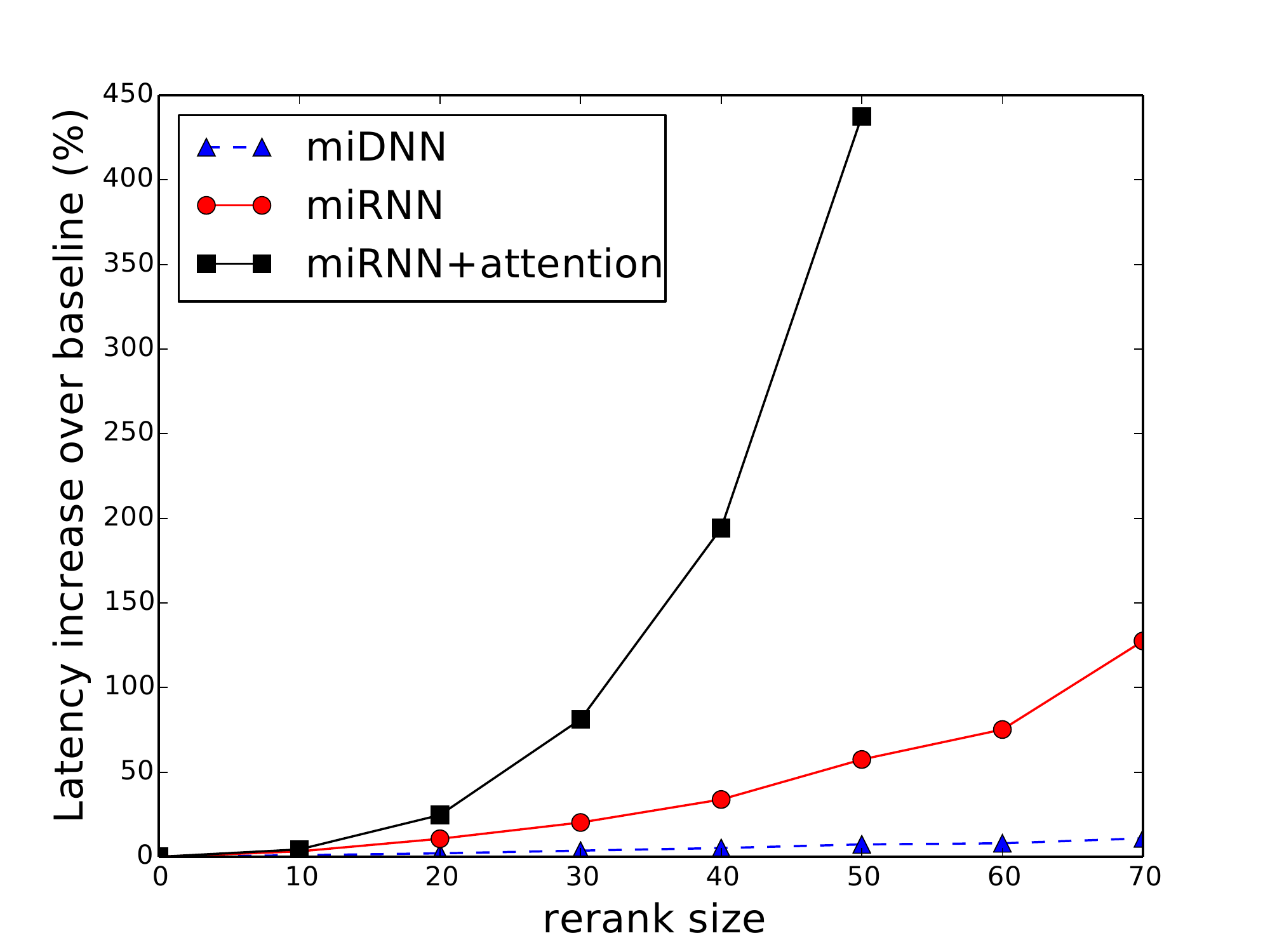}
    \caption{
        {\small The search latency increase with respect to rerank size.}
    \label{fig:latency_reranksize}}
\end{figure}

Figure~\ref{fig:gmv_reranksize} shows that our models increase GMV significantly over the baseline.
The GMVs of our models increase as rerank size grows, but gradually stabilize as rerank size gets larger than 50.
This may be explained by the statistics that 82\% users browses no more than 50 items.
So the benefit of increasing rerank size gradually gets small.
Note that the maximum rerank size of miRNN+attention is limited to 50 for computing efficiency.
To study the additional computational cost of our models, we compare the online search latency of different models.
The latency of the baseline is 21 ms. 
And the relative latency increase of our models over the baseline are shown in Figure~\ref{fig:latency_reranksize}.

In Figure~\ref{fig:latency_reranksize}, the latency of miDNN is small and grows linearly with respect to rerank size.
But the latency of miRNN and miRNN+attention grows polynomially.
When rerank size is 50, the latency of miRNN+attentions increases 400\% over the baseline, from 21 ms to 105 ms.
Although the RNN models achieves larger GMV, the computational cost of the RNN models are huge when rerank size gets big.
The large computational cost is the major drawback of our RNN models.

\begin{figure}[!tbh]
    \centering
    \includegraphics[width=0.34\textwidth]{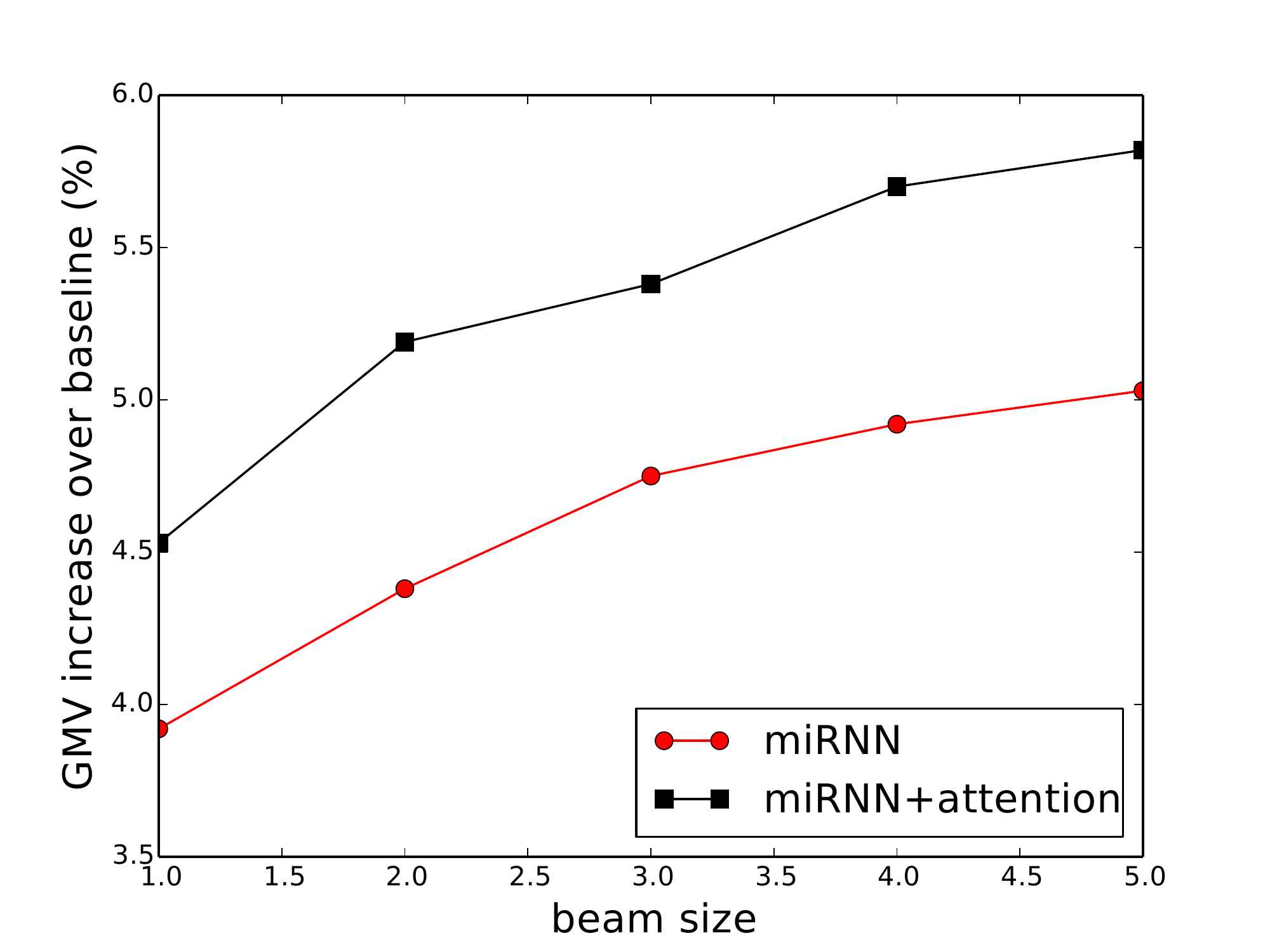}
    \caption{
        {\small The GMV increase with respect to beam size.}
    \label{fig:gmv_beamsize}}
\end{figure}

\begin{figure}[!tbh]
    \centering
    \includegraphics[width=0.34\textwidth]{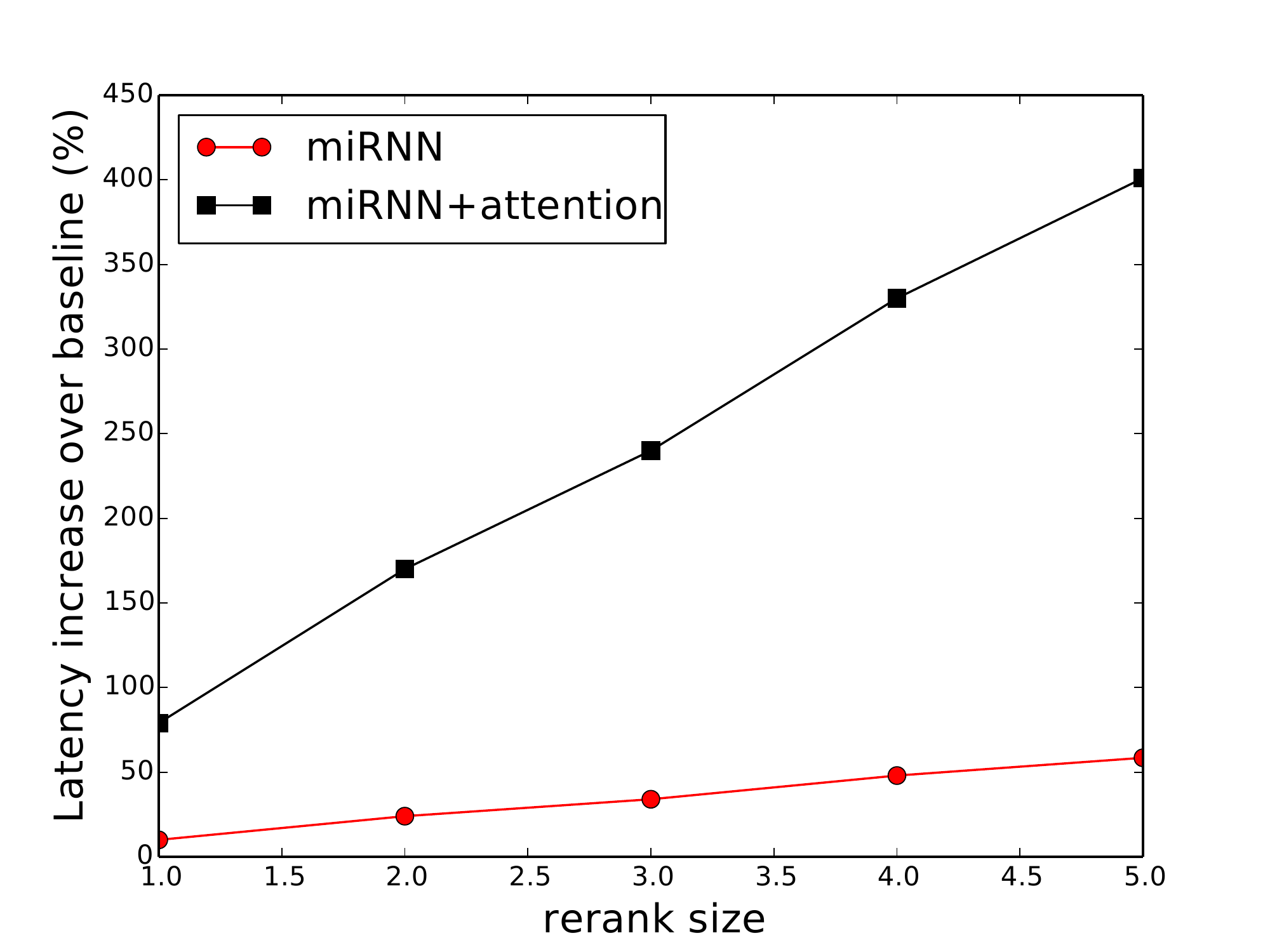}
    \caption{
        {\small The search latency increase with respect to beam size.}
    \label{fig:latency_beamsize}}
\end{figure}

For RNN models, we use beam search to find a good ranking sequence.
The beam size $k$ is a key parameter for beam search.
Larger $k$ means larger search space and usually results in better ranking results.
But larger $k$ also lead to more computational cost.
We studied the GMV and latency increase with respect to beam size.
And the results are shown in Figure~\ref{fig:gmv_beamsize} and Figure~\ref{fig:latency_beamsize}.

Figure~\ref{fig:gmv_beamsize} shows that the GMV increases as beam size grows.
But the GMV increase gets smaller when beam size gets larger.
Figure~\ref{fig:latency_beamsize} shows that the latency increases linearly with respect to beam size, which is in accordance with our time complexity analysis.
A balance of GMV and latency is needed to choose the value of beam size.
And we set the beam size to 5.
\begin{table}[!htb]
    \centering
    {\small
        \begin{tabular}{|c|c|c|c|c|}
            \hline
            Models & Rerank size & Beam size & GMV & Latency\\
            \hline
            miDNN & 50 & - & 2.91\% & 9\% \\
            miRNN & 50 & 5 & 5.03\% & 58\% \\
            miRNN+att. & 50 & 5 & 5.82\% & 401\% \\
            \hline
        \end{tabular}
    }
    \caption{The GMV increase in A/B test.\label{tab:gmv_latency_rst}}
\end{table}

Finally, we summarize our online test results in Table~\ref{tab:gmv_latency_rst}.
The rerank size is set to 50 and the beam size for RNN models is 5.
Results in Table~\ref{tab:gmv_latency_rst} show that our mutual influence aware ranking framework brings a significant GMV increase over the baseline.
The miDNN model achieves a good GMV increase with only a little latency overhead.
The miRNN+attention model gets the best GMV result but the latency grows too fast.
The miRNN model achieves a very good GMV increase with much less latency compared to miRNN+attention.
Therefore, if computational cost is very expensive, the miDNN model is a good choice.
In our case where mild latency increase is acceptable, the miRNN model is preferred.

\section{Conclusion}
In this paper, we point out the importance of mutual influences between items in e-commerce ranking and propose a global optimization framework for mutual influence aware ranking for the first time.
We incorporate mutual influences into our models by global feature extension and modeling ranking as a sequence generation problem.
We performed online experiments on a large e-commerce search engine.
To reduce computational cost, we use our methods as a reranking process on top of the baseline ranking.
The results show that our method produces a significant GMV increase over the baseline, and therefore verifies the importance of mutual influences between items.
We also compared the computational costs of our methods.
Our miDNN model noticeably increases GMV without much computational cost.
Our attention mechanism for RNN model gets the best GMV result. 
But the computational cost of our attention mechanism is too high.
Future research will be focused on more efficient attention mechanisms that increase GMV with less computations.
\section*{Acknowledgments}
This work receives great help from our colleague Xiaoyi Zeng. We would also like to thank Xin Li and the Taobao Search Engineering team for helpful discussions and the system engineering efforts.

\bibliographystyle{named}
\bibliography{global_mutual_influence_rank}

\begin{thebibliography}{}

\bibitem[\protect\citeauthoryear{Agrawal \bgroup \em et al.\egroup
  }{2009}]{Agrawal:09}
Rakesh Agrawal, Sreenivas Gollapudi, Alan Halverson, and Samuel Ieong.
\newblock Diversifying search results.
\newblock In {\em International Conference on Web Search and Data Mining
  (WSDM)}. Association for Computing Machinery, Inc., February 2009.

\bibitem[\protect\citeauthoryear{Bahdanau \bgroup \em et al.\egroup
  }{2014}]{bahdanau:14}
Dzmitry Bahdanau, Kyunghyun Cho, and Yoshua Bengio.
\newblock Neural machine translation by jointly learning to align and
  translate.
\newblock {\em arXiv e-prints}, abs/1409.0473, September 2014.

\bibitem[\protect\citeauthoryear{Burges \bgroup \em et al.\egroup
  }{2007}]{Burges:07}
Christopher~J. Burges, Robert Ragno, and Quoc~V. Le.
\newblock Learning to rank with nonsmooth cost functions.
\newblock In B.~Sch\"{o}lkopf, J.~C. Platt, and T.~Hoffman, editors, {\em
  Advances in Neural Information Processing Systems 19}, pages 193--200. MIT
  Press, 2007.

\bibitem[\protect\citeauthoryear{Cao \bgroup \em et al.\egroup }{2007}]{Cao:07}
Zhe Cao, Tao Qin, Tie-Yan Liu, Ming-Feng Tsai, and Hang Li.
\newblock Learning to rank: From pairwise approach to listwise approach.
\newblock In {\em Proceedings of ICML 2007}, pages 129--136, New York, NY, USA,
  2007. ACM.

\bibitem[\protect\citeauthoryear{Carbonell and Goldstein}{1998}]{Carbonell:98}
Jaime Carbonell and Jade Goldstein.
\newblock The use of mmr, diversity-based reranking for reordering documents
  and producing summaries.
\newblock In {\em Proceedings of SIGIR 1998}, pages 335--336, New York, NY,
  USA, 1998. ACM.

\bibitem[\protect\citeauthoryear{Chapelle and Chang}{2011}]{Chapelle:11a}
O.~Chapelle and Y.~Chang.
\newblock Yahoo! learning to rank challenge overview.
\newblock In Olivier Chapelle, Yi~Chang, and Tie-Yan Liu, editors, {\em
  Proceedings of the Learning to Rank Challenge}, volume~14 of {\em Proceedings
  of Machine Learning Research}, pages 1--24, Haifa, Israel, 25 Jun 2011. PMLR.

\bibitem[\protect\citeauthoryear{Chen and Yan}{2012}]{Chen:12}
Ye~Chen and Tak~W. Yan.
\newblock Position-normalized click prediction in search advertising.
\newblock In {\em Proceedings of the 18th ACM SIGKDD International Conference
  on Knowledge Discovery and Data Mining}, KDD '12, pages 795--803, New York,
  NY, USA, 2012. ACM.

\bibitem[\protect\citeauthoryear{Cossock and Zhang}{2008}]{Cossock:08}
David Cossock and Tong Zhang.
\newblock Statistical analysis of bayes optimal subset ranking.
\newblock {\em IEEE TRANSACTIONS ON INFORMATION THEORY}, 54(11):5140–--5154,
  2008.

\bibitem[\protect\citeauthoryear{Craswell \bgroup \em et al.\egroup
  }{2008}]{Craswell:08}
Nick Craswell, Onno Zoeter, Michael Taylor, and Bill Ramsey.
\newblock An experimental comparison of click position-bias models.
\newblock In {\em Proceedings of the 2008 International Conference on Web
  Search and Data Mining}, WSDM '08, pages 87--94, New York, NY, USA, 2008.
  ACM.

\bibitem[\protect\citeauthoryear{Joachims}{2002}]{Joachims:02}
Thorsten Joachims.
\newblock Optimizing search engines using clickthrough data.
\newblock In {\em Proceedings of the Eighth ACM SIGKDD International Conference
  on Knowledge Discovery and Data Mining}, KDD '02, pages 133--142, New York,
  NY, USA, 2002. ACM.

\bibitem[\protect\citeauthoryear{Li \bgroup \em et al.\egroup }{2008}]{Li:08}
Ping Li, Qiang Wu, and Christopher~J. Burges.
\newblock Mcrank: Learning to rank using multiple classification and gradient
  boosting.
\newblock In J.~C. Platt, D.~Koller, Y.~Singer, and S.~T. Roweis, editors, {\em
  Advances in Neural Information Processing Systems 20}, pages 897--904. Curran
  Associates, Inc., 2008.

\bibitem[\protect\citeauthoryear{Liu}{2009}]{Liu:09}
Tie-Yan Liu.
\newblock Learning to rank for information retrieval.
\newblock {\em Found. Trends Inf. Retr.}, 3(3):225--331, March 2009.

\bibitem[\protect\citeauthoryear{Radlinski \bgroup \em et al.\egroup
  }{2008}]{Radlinski:08}
Filip Radlinski, Robert Kleinberg, and Thorsten Joachims.
\newblock Learning diverse rankings with multi-armed bandits.
\newblock In {\em Proceedings of ICML 2008}, pages 784--791, New York, NY, USA,
  2008. ACM.

\bibitem[\protect\citeauthoryear{Sutskever \bgroup \em et al.\egroup
  }{2014}]{Sutskever:14}
Ilya Sutskever, Oriol Vinyals, and Quoc~V Le.
\newblock Sequence to sequence learning with neural networks.
\newblock In Z.~Ghahramani, M.~Welling, C.~Cortes, N.~D. Lawrence, and K.~Q.
  Weinberger, editors, {\em Advances in Neural Information Processing Systems
  27}, pages 3104--3112. Curran Associates, Inc., 2014.

\bibitem[\protect\citeauthoryear{Taylor \bgroup \em et al.\egroup
  }{2008}]{Taylor:08}
Michael Taylor, John Guiver, Stephen Robertson, and Tom Minka.
\newblock Softrank: Optimizing non-smooth rank metrics.
\newblock In {\em Proceedings of the 2008 International Conference on Web
  Search and Data Mining}, WSDM '08, pages 77--86, New York, NY, USA, 2008.
  ACM.

\bibitem[\protect\citeauthoryear{Wang \bgroup \em et al.\egroup
  }{2016}]{Wang:16}
Yue Wang, Dawei Yin, Luo Jie, Pengyuan Wang, Makoto Yamada, Yi~Chang, and
  Qiaozhu Mei.
\newblock Beyond ranking: Optimizing whole-page presentation.
\newblock In {\em Proceedings of the Ninth ACM International Conference on Web
  Search and Data Mining}, WSDM '16, pages 103--112, New York, NY, USA, 2016.
  ACM.

\bibitem[\protect\citeauthoryear{Xia \bgroup \em et al.\egroup }{2008}]{Xia:08}
Fen Xia, Tie-Yan Liu, Jue Wang, Wensheng Zhang, and Hang Li.
\newblock Listwise approach to learning to rank: Theory and algorithm.
\newblock In {\em Proceedings of ICML 2008}, pages 1192--1199, New York, NY,
  USA, 2008. ACM.

\bibitem[\protect\citeauthoryear{Xia \bgroup \em et al.\egroup }{2017}]{Xia:17}
Long Xia, Jun Xu, Yanyan Lan, Jiafeng Guo, Wei Zeng, and Xueqi Cheng.
\newblock Adapting markov decision process for search result diversification.
\newblock In {\em Proceedings of the 40th International ACM SIGIR Conference on
  Research and Development in Information Retrieval}, SIGIR '17, pages
  535--544, New York, NY, USA, 2017. ACM.

\bibitem[\protect\citeauthoryear{Xu and Li}{2007}]{Xu:07}
Jun Xu and Hang Li.
\newblock Adarank: A boosting algorithm for information retrieval.
\newblock In {\em Proceedings of the 30th Annual International ACM SIGIR
  Conference on Research and Development in Information Retrieval}, SIGIR '07,
  pages 391--398, New York, NY, USA, 2007. ACM.

\bibitem[\protect\citeauthoryear{Zhai \bgroup \em et al.\egroup
  }{2003}]{Zhai:03}
Cheng~Xiang Zhai, William~W. Cohen, and John Lafferty.
\newblock Beyond independent relevance: Methods and evaluation metrics for
  subtopic retrieval.
\newblock In {\em Proceedings of the 26th Annual International ACM SIGIR
  Conference on Research and Development in Informaion Retrieval}, SIGIR '03,
  pages 10--17, New York, NY, USA, 2003. ACM.

\bibitem[\protect\citeauthoryear{Zhu \bgroup \em et al.\egroup }{2014}]{Zhu:14}
Yadong Zhu, Yanyan Lan, Jiafeng Guo, Xueqi Cheng, and Shuzi Niu.
\newblock Learning for search result diversification.
\newblock In {\em Proceedings of SIGIR 2014}, pages 293--302, New York, NY,
  USA, 2014. ACM.

\end{thebibliography}

\end{document}